\documentclass[a4paper,fleqn,usenatbib]{mnras}

\usepackage[T1]{fontenc}
\usepackage{ae,aecompl}

\usepackage{graphicx}	
\usepackage{amsmath}	
\usepackage{amssymb}	

\title[Solar luminosity bounds on mirror matter]{Solar luminosity bounds on mirror matter}

\author[E. Michaely et al.]{
Erez Michaely$^{1}$\thanks{E-mail: erezmichaely@gmail.com},
Itzhak Goldman$^{2,3},$
and Shmuel Nussinov$^{2}$
\\
$^{1}$Astronomy Department, University of Maryland, College Park, MD \\
$^{2}$School of Physics and Astronomy, Tel-Aviv University, Tel-Aviv, Israel\\
$^{3}$Department of Physics, Afeka Tel-Aviv Engineering College, Tel-Aviv, Israel
}

\date{Accepted XXX. Received YYY; in original form ZZZ}

\pubyear{2019}

\begin{document}
\label{firstpage}
\pagerange{\pageref{firstpage}--\pageref{lastpage}}
\maketitle

\begin{abstract}
We present bounds on mirror dark matter scenario derived by using
the effect of mirror matter on the luminosity of the Sun. In the perturbative
regime where the mirror matter concentration is small relative to
the ordinary matter we estimate the heat transfer from ordinary matter
to the mirror sector by simple analytic consideration. That amount
of heat transfer is radiated via mirror photons and increases the
required energy production in order to maintain the observed luminosity.
We then present more detailed numerical calculations of the total
amount of this energy transfer.
\end{abstract}




\section{Introduction}

Finding the nature of Dark Matter (DM) which most likely contributes
$\sim25\%$ of the energy density of the universe is an outstanding
challenge \cite{Feng2010}. Some types of DM which arise beyond the
standard model (BSM) of particle physics are being experimentally
searched and or are constrained by astrophysics. In particular DM
particles such as axions, dark photons and new types of neutrinos
can be emitted from and lead to excessive cooling of neutron stars,
white dwarfs red giants etc. \cite{Raffelt1996}.

The apparent inconsistency of the observed solar neutrino fluxes with
the predicted values was the first indication \cite{Bahcall1989,Bahcall2004,Bahcall2006}
for the only piece of BSM physics (apart from DM!) known today, namely
that of neutrino masses and mixings\cite{Tanabashi2018}. Early on
it motivated an alternative suggestion that small accumulation in
the sun of weakly interacting DM particles (WIMPs) could cool the
solar core and evade the \textquotedbl solar neutrino problem\textquotedbl{}
\cite{Spergel1985,Press1985,Faulkner1985,Gilliland1986}. Indeed at
that time the main difficulty was the paucity of the energetic $\ ^{8}B$
neutrinos whose rate scales as $T_{0}^{24}$ with $T_{0}$ is the
central temperature.

In this note we again use the Sun to limit DM arising in mirror models
in which a hidden sector exists where every particle or parameter
in the standart model - $x$ is mirrored by an identical particle/
parameter $x'$ \cite{Lee1956,Foot2014,Kobzarev1968}; for review
see \cite{Foot2014} and references within. In much of the work these
models the symmetry between the mirror and ordinary sector is broken
allowing different masses of particles and their mirrors which helped
address many astrophysical and cosmological issues \cite{Berezhiani1996}.
Still almost exact mirror models with minor changes made to allow
$\Omega\left(B'\right)=5\Omega\left(B\right)\sim\Omega\left({\rm DM}\right)$
and a lower CMB temperature in the mirror sector (which is required
to avoid conflict with BBN- Big Bang Nucleosynthesis) have been discussed
at length \cite{Berezhiani2001,Foot2014}. It was suggested that this
most restrictive framework with appropriate minimal weak mirror-ordinary
matter interaction, can evade the bullet cluster bound and the infall
of DM into parallel galactic disks expected for the strongly mutually
interacting and dissipative DM made up of mirror atoms. Here we will
focus on the possible effect of the accumulation of mirror DM particles
arising in the framework of \textquotedbl almost exact\textquotedbl{}
mirror symmetry in the Sun at a relative concentration
\begin{equation}
\eta=\frac{M'}{M_{\odot}}
\end{equation}
where $M'$ is the overall mass of the mirror particles in the Sun.
We find that the resulting changes of the solar luminosity exclude
$\eta$ and mirror-ordinary matter cross-sections $\sigma_{{\rm xx'}}$
values in a region allowed by all other constraints \textit{and} which
was strongly favoured by mirror DM models. This exclude most exact
mirror DM variants.

\section{The effect of mirror matter accumulation in the Sun}

We start by describing the differences between earlier attempts to
constrain massive DM particles by considering the consequences of
their accumulation in the Sun and our present discussion of almost
exact mirror model particles.

In the earliest works mentioned above \cite{Press1985,Spergel1985},
only the mutual DM-nuclear cross-section $\sigma_{{\rm xx'}}$ are
used to trap in the sun DM particles of masses in the $5-10{\rm GeV}$
range. In this case $\sigma_{{\rm xx'}}\gtrsim10{}^{-36}{\rm cm^{2}}$
was required in order to accumulate the minimal $\eta\sim10^{-11}$
which allows sufficient heat transport from the solar core so as to
significantly reduce the $\ ^{8}B$ neutrino flux. Extensive direct
searches for DM in large cryogenic, underground detectors restrict
by now \cite{Tanabashi2018} the He' (the dominant mirror dark matter
component in mirror models) nucleon cross-section, to be less than
$10^{-38}{\rm cm^{2}}$ \cite{Foot2014}.

The next class of DM particles considered in this context was that
of asymmetric, strongly self interacting DM \cite{Frandsen2010,Cumberbatch2010,Taoso2010}.
In this case newly falling DM particles can be captured by scattering
on DM particles which were captured earlier in the Sun. This increases
the capture rate until it reaches the \textquotedbl Unitarity bound\textquotedbl{}
when essentially every DM particle hitting the Sun is captured. The
integrated accumulation over the solar lifetime can then lead to the
concentration of $\eta\sim10^{-11}$, the value mentioned above. With
the present understanding of neutrino mixing, extra heat convection
from the very central region is no longer required to explain the
solar neutrino \textquotedbl Problem\textquotedbl . The resulting
cooling of the core still has other more subtle yet observable effects
on the standard solar model as discussed in \cite{Frandsen2010,Cumberbatch2010,Taoso2010}.

The strong mutual x'-x' scattering helps retain the captured DM inside
the Sun. Indeed in the absence of such strong scattering some x's
with energy of $E\sim{\rm few}\ kT_{0}\sim{\rm few}\ {\rm KeV}$ and
a velocity $v=\left(2E/m_{{\rm x'}}\right)^{1/2}$ exceeding the escape
velocity from the solar core of $\sim1000{\rm Km/Sec}$ will be kicked
from the solar core, if the DM particles are lighter than $\sim5{\rm GeV}$.
This limit does not apply for strongly self interacting DM: the kicked
x' suffers many collisions with the ambient x's quickly sharing its
energy and no escape of DM particles is expected even if $m_{{\rm x'}}=m_{H}\approx{\rm GeV}$.

The Mirror dark matter considered here differs from that in the above
two cases. Thanks to the exchange of the mass-less mirror photon it
is strongly interacting via Ruthdeford scattering

\begin{equation}
\sigma_{{\rm x'x'}}\sim\alpha^{2}/E^{2}=\sigma_{{\rm xx}}
\end{equation}

which is $\sim10^{-18}{\rm cm^{2}}$ for the relevant KeV energies.
\footnote{In the solar core the plasma (Debey) screening correction to the above
cross-section amounts to replacing the momentum transfer squared $k^{2}$
in the photon momentum space propagator by $k^{2}\rightarrow k^{2}+k_{s}^{2}$
where $k_{s}^{2}=k_{D}^{2}=4\pi\alpha n\left(e\right)/T\sim40{\rm KeV^{2}}$.
Using the temperature and electron density appropriate to the solar
core we find $k_{s}^{2}\approx40{\rm KeV^{2}}$. This screening cuts-off
the very forward, low momentum transfer scattering but only mildly
affects the relevant transport cross-section as the average $k^{2}\sim2m\Delta E\sim2mT\sim40{\rm KeV^{2}}$
even for the case of e-e or e'-e scattering with $m\approx1/2{\rm MeV}$.
The net effect of replacing $1/k^{2}$ by $1/\text{\text{\ensuremath{\left(k^{2}+k_{s}^{2}\right)}}}$
is to reduce the cross-sections by just a factor of two \cite{Raffelt1996}.}

The other most important feature is that mirror nuclei/electrons can
emit the massless mirror photons and therefore constitute dissipative
DM.

To make our argument as model independent as possible we use the concentration
of the mirror particles in the Sun $\eta$ and the ordinary- mirror
scattering cross-section $\sigma_{{\rm xx'}}$ as the two independent
parameters of the particle physics model to be constrained by the
astrophysical considerations. In generic almost exact mirror models
not only the cross-sections for mirror- mirror interactions are fixed
by the above Ruthdeford scattering but also the coupling of mirror
charged particles with the ordinary sector is via the \textquotedbl Photon
Portal\textquotedbl{} - namely the kinetic mixing: $\epsilon F^{\mu,\nu}F'_{\mu,\nu}$
of the mirror photon field strength tensor $F'_{\mu,\nu}=\partial_{\nu}A'_{\mu}-\partial_{\mu}A'_{\nu}$
and the field strength tensor of the ordinary photon. Since both our
photon $\gamma$ and the mirror photon $\gamma'$ are massless we
should identify our physical photon $\tilde{A}$ with the superposition
of original fields: $\tilde{A}_{\mu}=A_{\mu}+\epsilon A_{\mu}^{'}$.
Indeed the two fields $A$ and $A'$ in this particular superposition
are coherently emitted, propagated and absorbed by SM charges. This
redefinition then subsumes \textit{all} the ordinary matter mirror
photon interaction and therefore ordinary particles no longer couple
to the mirror photon. Yet each mirror particle $x'$ with a mirror
electric charge of $e'\left(x'\right)=e\left(x\right)\equiv e$ couples
to the ordinary photon with a milli-charge of $\epsilon e$. The cross-section
for mirror- ordinary matter x'-x scattering generated by ordinary
photon exchange is then the standard Ruthdeford scattering above reduced
by $\epsilon^{2}$
\begin{equation}
\sigma_{{\rm x'x}}=\epsilon^{2}\sigma_{{\rm xx}}\approx\epsilon^{2}\frac{\alpha^{2}}{E^{2}}.\label{eq:sigma_x_xprime}
\end{equation}

To address the apparent departure from exact mirror symmetry we note
that the above definition is appropriate in regions of space which
are dominated by ordinary matter. The opposite scheme where the redefined
physical mirror photon does not couple to ordinary matter and ordinary
charged particles are milli-charged with respect to $A'$ is appropriate
in regions dominated by mirror matter such as the interior of mirror
stars discussed later.

The effect of capturing mirror particles differs from that in the
previous case due to the fact that the mirror particles radiate dark
(mirror) photons. This provides yet another channel for radiating
the energy generated by nuclear reactions in the solar core- a channel
which operates in parallel with the usual ordinary photon radiation.
Thus the mirror matter not only transports heat, but just like the
emission of axions or massive dark photons that directly couple to
the solar nuclei/ electrons , it also changes the overall solar energetics.
This will allow us to derive limits on the mirror concentration $\eta$
and the mirror ordinary sector particle scattering cross-section $\sigma_{{\rm xx'}}$
which are more robust and less model dependent than earlier limits.

A second important difference between the almost exact mirror matter
discussed here and the previous merely strongly interacting DM, is
that it's concentration in the Sun is \textit{no longer} restricted
by the maximal capture rate during the lifetime of the Sun to be:
$\eta\sim10^{-11}$.

Both ordinary and mirror matter are disipative and mutually attract
gravitationally. We therefore expect ordinary matter to cluster in
the gravitational wells of mini haloes generated in the mirror matter.
Conversely, mirror matter should cluster in the gravitational wells
due to ordinary matter galactic disks. This co-clustering or even
co-collapses tends to mix the two types of matter. In particular it
could lead to an initial mirror matter concentration in the Sun which
much exceeds the above limiting $\eta\sim10^{-11}$. Indeed it has
been estimated \cite{Foot2014} that the original pre-solar cloud
can efficiently accumulate mirror particles leading to $\eta\sim10^{-5}$.

It is important to note that mirror matter is still a rather subdominant
component of the Sun. This justifies treating the effect of this admixture
perturbativly using the known solar profiles of ordinary density,
$\rho\left(r\right)$, and temperature, $T\left(r\right)$.

For solar/stellar cooling by weakly interacting particles emitted
by nucleons and/or electrons it suffices to compute the volume emission
of these particles which freely stream out. This is not the case here.
First the mirror photons are not emitted directly from the electron/protons
 in the Sun but only by the mirror particles after kinetic/heat
 energy is transferred to them via collisions with ordinary core
particles at a total rate which we denote by $dQ/dt$. Thanks to their
very strong mutual interactions the mirror particles will then equilibrate
and generate at each radius a local temperature profile $T'\left(r\right)$.
In general this $T'\left(r\right)$ is different from $T\left(r\right)$,
the temperature profile of ordinary protons/ electrons. These mirror
particles will then emit their energy via mirror photons generated
by bremsstrahlung in $e'-p'$ or $e'-\alpha'$ collisions. Also unlike
for the simple volume emission case mentioned above, the mirror photons
will often scatter on their way out on the ambient mirror particles
and will be trapped for some time $\tau'$. Both $dQ/dt$ and $\tau'$
depend on the, as yet unknown, profiles of the density $\rho'\left(r\right)\sim n'\left(r\right)$
and of the temperature $T'\left(r\right)$ of the mirror matter. Before
embarking on the calcuation of these profiles and the resulting $dQ/dt$
we will first make some estimates using a simpler approach.

\subsection{\label{subsec:Estimate-of-the}Estimate of the energy transfered
and radiated by the mirror photons}

While we find that $dQ/dt$ is comparable with the observed ordinary
solar luminosity $L_{\odot}$ for a range of allowed DM parameters
$\sigma_{{\rm xx'}}$ and $\eta$, the reverse heat flow to the original
matter reservoir $dQ'/dt$ is \textit{negligible}. The reason for
this are the large self scattering of mirror matter $\sigma_{{\rm x'x'}}$
which exceeds by $\epsilon^{-2}>10^{18}$ the mirror-ordinary particle
collision cross-section $\sigma_{{\rm xx'}}$ and the large bremsstrahlung
cross-section leading to $\gamma'$ emission: $\sigma_{{\rm x'x'}\rightarrow{\rm x'x}'+\gamma'}\sim\alpha\times\sigma_{{\rm x'x'}\rightarrow{\rm x'x'}}\sim10^{-20}{\rm cm^{2}}$.
Thus a mirror particle which has gained energy by a collision with
an ordinary particle in the solar core will collide and share it's
energy with other ambient mirror particles rather than collide again
with an ordinary ion or electron. In turn bremsstrahlung quickly transfer
this energy to mirror photons. Since these mirror photons do not scatter
at all from protons or electrons but only from mirror particles the
energy transferred to the mirror sector, be it the matter or radiation
part, stays in that sector and eventually is emitted as mirror photons.
Thus to find the extra luminosity emitted via mirror photons we need
only to find $dQ/dt$.

We will mainly focus on the inner core, $R\approx0.2R_{\odot}=1.4\times10^{10}{\rm cm}$
which includes a total mass $M\left(R\right)\approx0.35M_{\odot}$
and generates $\sim90\%$ of the solar luminosity \cite{Paxton2011}.
The mainly ordinary matter densities therein of $\sim165{\rm gr/cm^{3}}$
corresponds to electron number density of $n_{e}\approx4.8\cdot10^{25}{\rm cm^{-3}}$.
The almost constant temperature is on average $T\sim1.3{\rm KeV}$
or $2\cdot10^{-9}{\rm ergs/particle}$. Since the Ruthdeford scattering
depends only on the energy and not the mass of the colliding particles
and in thermal equilibrium electrons have the same energy of $3/2kT$
as protons or He ions, the scattering of the faster moving $e$ and
$e'$ will dominate the heat transfer process. Such scattering of
two equal mass particals tends to equalise their energy and on average
an energy 
\begin{equation}
\Delta E=3/4\left(kT\left(r\right)-kT'\left(r\right)\right)\approx\frac{3}{4}kT\left(r\right)=1.5\times10^{-9}{\rm erg}
\end{equation}
will be transferred to the mirror sector in each collision. The density
profile of mirror particles is given by the Boltzmann distribution
$\propto exp{\left(-V(r)/{kT'}\right)}$. $V\left(r\right)\approx4\pi/3G\rho_{0}m_{{\rm He}}r^{2}$
is the gravitational potential due to the ordinary roughly constant
density $\rho_{0}=165{\rm gr/cm^{3}}$ of ordinary matter. The mass
$m_{{\rm He'}}$ rather than $m_{{\rm p'}}$ was used as mirror helium
is the dominant component in the mirror sector, namely 
\begin{equation}
X'\equiv\frac{\rho'_{H'}}{\rho'}=0.2\ \text{and }Y'\equiv\frac{\rho'_{{\rm He}}}{\rho'}=0.8.
\end{equation}
\cite{Foot2014}. $\rho'\left(r\right)$ then is a Gaussian, $\exp{-(r/r_{0})^{2}}$
with 
\begin{equation}
r_{0}=\left(\frac{4\pi}{3}\frac{kT'}{G\rho_{0}m_{{\rm He}}}\right)^{1/2}\approx10^{10}{\rm cm}.
\end{equation}
Since this is less than $R\approx0.2R_{\odot}$ most mirror particles
are inside the above core. Each mirror electorn experiences 
\begin{equation}
\Gamma_{ee'}=n_{e}v_{e}\sigma_{ee'}\approx1.2\times10^{35}\sigma_{ee'}{\rm sec}^{-1}
\end{equation}
 collisions per second, where $v_{e}=\left(3kT/m_{e}\right)^{1/2}$
is the electron velocity and $n_{e}\approx4.8\cdot10^{25}{\rm cm^{-3}}$
is the electron number density.

The total energy transferred to the mirror particles per second then
is: 
\begin{equation}
\frac{dQ}{dt}=N_{{\rm tot,e'}}\Gamma_{ee'}\Delta E
\end{equation}
where 
\begin{equation}
N_{{\rm tot,e'}}=N_{H'}+2N_{{\rm He'}}=\frac{\eta M_{\odot}}{m_{H}}\left(X'+\frac{1}{2}Y'\right)=0.6\frac{\eta M_{\odot}}{m_{H}}\approx\eta7.2\times10^{56}.
\end{equation}
Hence the total energy transfer per second is 
\begin{equation}
\frac{dQ}{dt}=\eta\sigma_{{\rm ee'}}1.3\times10^{83}\frac{erg}{sec}{\rm =\eta\sigma_{ee'}3.4\times10^{49}L_{\odot}}.
\end{equation}
Rewriting the previous equation in terms of $\sigma_{-38}\equiv\sigma_{ee'}/10^{-38}$
and $\eta_{-11}\equiv\eta/10^{-11}$ we find 
\begin{equation}
\frac{dQ}{dt}=\eta_{-11}\sigma_{-38}3.4L_{\odot.}
\end{equation}

The rather modest requirement that $dQ/dt$, the mirror photon luminosity
will not exceed the $0.04L_{\odot}$ then limits the region of allowed
parameters by: 
\begin{equation}
\eta_{-11}\sigma_{-38}<1.1\times10^{-2}.
\end{equation}

The rational for requiring $dQ/dt<0.04L_{\odot}$ is that the flux
of $pp$ solar neutrinoes which directrly reflects the nuclear energy
output is measured and understood at $\sim4\%$ level \cite{Bergstroem2016}.
Originally considerations of energy loss were used e.g. by \cite{Raffelt1996} in a conservative way, requiring only that the
new extra luminosity will not exceed the ordinary luminosity. During
the past decades measurements of all types (pp, Berilium, Boron ,
etc) of solar neutrinos and the understanding of their apparent deficit
via neutrino mixing, the parameters of which was independently measured
in terrestrial experiments, have greatly improved \cite{Vissani2017}.

\begin{figure*}
\includegraphics[width=0.75\paperwidth]{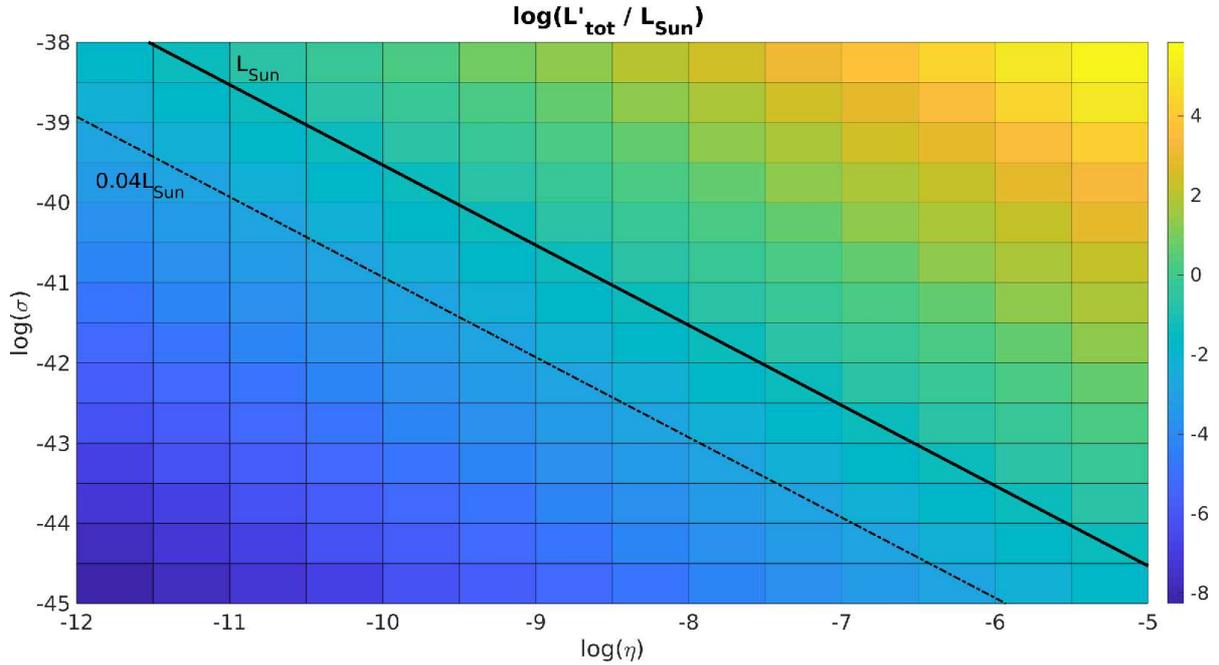}\caption{\label{fig:grid}The calculated $\log\left(L'_{{\rm tot}}/L_{\odot}\right)$
for the 15X15 different values of $\eta$ and $\sigma_{{\rm xx'}}$.
For each pair of $\left(\eta,\sigma_{{\rm xx'}}\right)$ we found
the density and temperature profile of the mirror particles (see next Sec. for details) and calculated the total
energy transferred, $L'_{{\rm tot}}$. The black solid line corresponds
to the estimated (subsection \ref{subsec:Estimate-of-the}) energy transfer
of $1L_{\odot}$ and the dot-dashed line corresponds for $0.04L_{\odot}.$
Any pair of $\left(\eta,\sigma_{{\rm xx'}}\right)$ above this line
is excluded.}
\end{figure*}

\section{\label{sec:Numerical}Numerical calculation of the mirror
luminosity in the Sun}

In this section we describe the numerical results and calculation
of $dQ/dt$, the rate of the total heat transferred from the ordinary
matter to the mirror matter. The amount of heat transferred to the
mirror particles depends on their density profile (number density),
$\rho'\left(r\right)$ $\left(n'\left(r\right)\right)$ and temperature
profile, $T'\left(r\right)$. For a given $\eta$ and $\sigma_{{\rm ee'}}$
these functions are unknown apriori, however they must satisfy the
four well known stellar structure equations:
\begin{equation}
\frac{dP'\left(r\right)}{dr}=-\frac{GM\left(r\right)\rho'\left(r\right)}{r^{2}}\label{eq:hydrostatic}
\end{equation}
\begin{equation}
\frac{dM'\left(r\right)}{dr}=4\pi r^{2}\rho'\left(r\right)\label{eq:mass_continuity}
\end{equation}
\begin{equation}
\frac{dT'\left(r\right)}{dr}=-\frac{3L'\left(r\right)\kappa'\left(r\right)\rho'\left(r\right)}{4\pi r^{2}4acT'\left(r\right)^{3}}\label{eq:radiative_transport}
\end{equation}
\begin{equation}
\frac{dL'\left(r\right)}{dr}=4\pi r^{2}\rho'\left(r\right)\mathcal{\epsilon}'\left(r\right)\label{eq:energy_conservation}
\end{equation}
where $\kappa'$ is the opacity for Thomson scattering, $a$ is the
radiation constant and $\mathcal{\epsilon}'$ is the energy source
per unit mirror mass of the mirror particles. We identify the product
$\rho'\left(r\right)\mathcal{\epsilon}'\left(r\right)$ to be the
transfers heat per unit mirror mass.

\begin{figure}

\includegraphics[width=1\columnwidth]{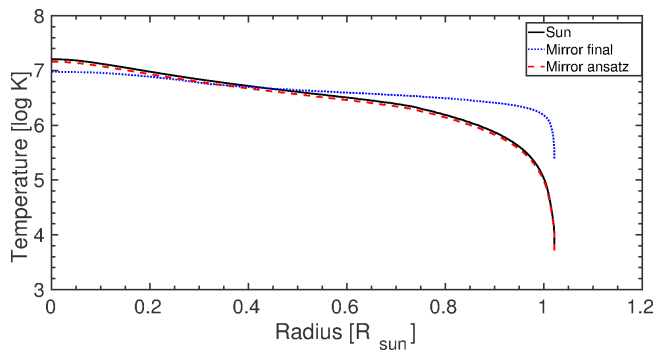}\caption{\label{fig:profiles}Black solid line, is the Sun temperature profile,
$T\left(r\right)$. Red dashed line is the initial ansatz, $T'\left(r\right)=0.9T\left(r\right)$.
Blue dotted line is the calculate profile that satisfies equation
(\ref{criteria}).}

\end{figure}

For a specific pair of $\eta$ and $\sigma_{{\rm ee'}}$ we postulate
an ansatz for $\rho'\left(r\right)=\eta\rho\left(r\right)$ and $T'\left(r\right)=0.9T\left(r\right)$,
where $\rho\left(r\right)$ and $T\left(r\right)$ are the density
and temperature profiles of the Sun taken from MESA, stellar evolution
code \cite{Paxton2011}. Using this ansatz one can calculate the left
hand side (lhs) and right hand side (rhs) of equations (\ref{eq:hydrostatic}-\ref{eq:energy_conservation}).
The ratios of the lhs and the rhs, $q_{i}$, where $i$ runs over
the above four stellar structure equations, is a measure of the quality
of the initial guess. We repeatedly altered the functions $\rho'\left(r\right)$
and $T'\left(r\right)$ in order to minimize $\left|q_{i}-1\right|$.
We are able to find the profiles that satisfy the stellar structure
equations within the tiny errors so that the computed integrated luminosity
satisfies 
\begin{equation}
\frac{\int L'_{{\rm n+1}}dr-\int L'_{{\rm n}}dr}{\int L'_{{\rm n}}dr}<5\times10^{-9}\label{criteria}
\end{equation}
where $n$ indicated that n-th iteration. In figure \ref{fig:profiles}
we present a representative example, the black solid line is the Sun
temperature profile, $T\left(r\right)$, while the red dashed line
is the ansatz, $T'\left(r\right)=0.9T\left(r\right)$. After many
iterations that minimize $q_{i}$, the caclulated profile satisfied
eq. (\ref{criteria}), we found the blue dotted line. The same mechanism
is done for the density profile, $\rho'\left(r\right)$.

Once we find the mirror density, $\rho'\left(r\right)$ and temperature
profile, $T'\left(r\right)$ for a pair of $\eta$ and $\sigma_{{\rm ee'}}$
we calculate the total energy transferred and hence emitted by mirror
photon and record it. We calculated $dQ/dt\equiv L'_{{\rm tot}}=\int L'dr$
for the following parameters: 15 values equally spaced in log of the
mirror matter- ordinary mirror cross-section $\sigma_{{\rm xx'}}=\left\{ 10^{-45}-10^{-38}\right\} $
and 15 values equally spaced in log of $\eta=\left\{ 10^{-12}-10^{-5}\right\} .$
Figure \ref{fig:grid} present the results of our calculation on the
above 15X15 grid. The results are presented in terms of $\log\left(L'_{{\rm tot}}/L_{\odot}\right)$.
Our results agree well with our estimate from subsection \ref{subsec:Estimate-of-the}
for $1L_{\odot}$ (black solid line) and $0.04L_{\odot}$ (black dot-dashed
line). Any pair of values $\left(\eta,\sigma_{{\rm xx'}}\right)$
which is above the $0.04L_{\odot}$ is therefore excluded.

\section{Discussion and summary}

We start by pointing that:
\begin{itemize}
\item Since our analysis above was essentially perturbative in nature it
cannot directly apply to the cases where the computed mirror photon
luminosity considerably exceeds the $4\%$ of the solar luminusity,
as the ordinary solar parameters would need then to be modified as
well. Still it is quite safe to assume that the very large mirror
luminosity arising when $\eta\sigma_{{\rm xx'}}>10^{-46}$ will be
indicative of some fatal difficulties with the observed Sun.
\item In order to evade the Bullet cluster upper bound on the dark dark
(here mirror-mirror scattering cross-section) we need that only some
fraction, say $15\%,$ of the mirror matter in the haloes will stay
unclussterd and that majority form collisonless stars. The argument
for $\eta\sim10^{-5}$ can then be \textquotedbl mirrored \textquotedbl{}
to suggest a similar admixture of ordinary matter within the mirror
stars. If the latter are still active then even a small fraction of
their nuclear energy production channeled into ordinary radiation,
analogous to that found above, will make theses stars visible and
no allow them to be DM in the first place.
\end{itemize}
To summarize we note that the most relevant difference between our
and previous limits steming from DM captured in the Sun is the fact
that we use the solar luminosity rather than more subtle aspects like
Helio-seismography. This in turn limits the ${\it {product}}$ of
$\eta$ the solar concentration of DM and $\sigma_{{\rm xx'}}$ the
dark-ordinary matter cross-sections, rather than each of these separately.
This is particularly relevant for the case of (almost) symmetric mirror
models which provided the framework of the present analysis. The point
is that in these models both $\sigma_{xx'}$ and $\eta$ are fixed
by the same single dimensionless kinetic mixing parameter $\epsilon$
of the photon and mirror photon. Specifically $\epsilon^{2}$ appears
in the mirror-ordinary matter Ruthdeford like scattering (\ref{eq:sigma_x_xprime})
above. In order to evade the apparent difficulties associated with
mirror matter forming a disc overlaping the ordinary Milky Way disc
one needs a minimal $\sigma_{{\rm xx'}}$ corresponding to a high
epsilon value of $\sim10^{-9}$. The parameter $\epsilon$ also controls
the expected fraction of mirror matter $\eta\sim\epsilon^{2}$ which
is mixed into the presolar cloud. This preferred optimal value yields
the $\eta\sim10^{-5}$ and $\sigma_{{\rm xx'}}\sim10^{-36}$. The
product $\eta\sigma_{{\rm xx'}}\sim10^{-41}$ will then exceed the
maximum value we found from the $4\%$ limit of the solar luminosity
emitted via mirror photons to be $\eta\sigma_{{\rm xx'}}\sim10^{-51}$
, by a factor of $10^{10}$! Thus our new limits tends to most strongly
exclude the above optimal $\epsilon$ value and the large class of
almost exactly symmetric mirror models which depend on it.

\section*{Acknowledgements}

EM thanks Nathan Roth, Richard Mushotzky and Coleman Miller for stimulating discussions.










\bsp	
\label{lastpage}
\end{document}